\documentclass[%
reprint,
superscriptaddress,
amsmath,amssymb,
aps,
pra,
]{revtex4-1}

\usepackage{graphicx}% Include figure files
\usepackage{dcolumn}% Align table columns on decimal point
\usepackage{bm}% bold math
\begin{document}
	
\preprint{APS/123-QED}
	
\title{Thouless pumping and Haldane-model topology in a driven extended SSH chain}% Force 

\author{Zhen Dong}
\affiliation{College of Physics and Electronic Engineering, Sichuan Normal University, Chengdu 610068, China}
\author{Xin-Lei Zhao}
\affiliation{College of Physics and Electronic Engineering, Sichuan Normal University, Chengdu 610068, China}
\author{Yi Zheng}
\email{Corresponding author: zhengyireal@sicnu.edu.cn}
\affiliation{College of Physics and Electronic Engineering, Sichuan Normal University, Chengdu 610068, China}
	
%\date{\today}
	
\begin{abstract}
Thouless pumping provides a powerful route for probing Chern topology through quantized motion in one-dimensional lattices. We investigate such a topological pump in an extended Su-Schrieffer-Heeger (SSH) model with tunable next-nearest-neighbor hoppings. A graphical representation in Bloch space is used to construct pumping schemes. In this framework, the Bloch vector traces a closed loop that evolves in time. The Chern number of the synthetic band is then determined by the motion of this loop. Quantum pumping phenomena are demonstrated for a filled band and for a bosonic wave packet driven by Bloch oscillations. We further show that the Haldane model can be mapped to the driven extended SSH chain by replacing a quasimomentum by the pumping time. The quantized charge pump indicates that the Chern number of a two-dimensional model can be read out from a one-dimensional pumping protocol, offering an accessible route for detecting higher-dimensional topology, such as in photonic and cold-atom platforms.
\end{abstract}
	
%Use showkeys class option if keyword
\maketitle
	
%\tableofcontents
	
\section{INTRODUCTION}\label{sec1} 
Topological phases of matter have fundamentally extended the theoretical framework of condensed matter physics beyond spontaneous symmetry breaking. Topological band theory connects the global geometry of Bloch states to quantized observables\cite{qi2011topological,xiao2010berry}. In two dimensions, the Chern number determines the Hall response and is exemplified by the Haldane model, which realizes a Chern insulator without net magnetic flux \cite{thouless1982quantized,haldane1988model}. Implementing and diagnosing such Chern bands nevertheless requires carefully engineered complex hoppings and access to their momentum-space geometry, as illustrated by cold-atom realizations of the Haldane model \cite{jotzu2014experimental}. Thouless pumping offers a dimensional counterpart to this problem. For a gapped one-dimensional (1D) Hamiltonian that is varied adiabatically and periodically, momentum and time form a synthetic two-dimensional torus. The charge transported in one cycle is then quantized by the Chern number \cite{thouless1983quantization,citro2023thouless, niu1984quantised,king1993theory, rice1982elementary,wang2013topological}. This relation converts a two-dimensional topological invariant into a real-space displacement and has enabled direct observations with ultracold fermions and bosons \cite{nakajima2016topological,lohse2016thouless,lu2016geometrical,schweizer2016spin, schouten2021rapid,minguzzi2022topological,privitera2018nonadiabatic,fedorova2020observation}.
	
The Su-Schrieffer-Heeger (SSH) model provides a minimal 1D setting in which band topology is represented by the winding of a Bloch vector \cite{su1979solitons,zak1989berry, su1980soliton}. The two dimerization patterns and particular boundary condition generate a quantized Zak phase and robust zero-energy edge modes closely related to Jackiw--Rebbi solitons. This reflects the physical consequence of chiral symmetry and non-trivial topology. Given these features, adding next-nearest-neighbor (NNN) hopping substantially enriches this setting. In particular, it has been shown that cyclic modulation of both nearest-neighbor and NNN hoppings produces nonzero Chern numbers in the space spanned by momentum and a modulation parameter, together with a phase diagram closely resembling that of the Haldane model \cite{li2014topological}. On the other hand, introducing the NNN hopping generally breaks chiral symmetry, leaving the Zak phase of an instantaneous Hamiltonian unquantized. This enables Thouless pumping by implementing a cyclic route which connects topological trivial and nontrivial limits. 
		
Identifying a topological pumping path does not by itself guarantee an accessible measurement. Conventional quantization assumes uniform occupation of an isolated band, which is natural for a filled fermionic band but restrictive for bosonic wave packets. Their quasimomentum distribution can be narrow and strongly dependent on state preparation. Bloch-oscillation-assisted pumping relaxes this requirement by using a weak lattice tilt to sweep the state across the Brillouin zone \cite{ke2020topological}. The resulting time-domain sampling can recover Chern-number transport from generic single-band initial states. This strategy has since been extended to correlated bosons \cite{liu2023correlated}, while nonlinear soliton pumps and returning pumps have further broadened the scope of quantized topological transport \cite{mostaan2022quantized,cheng2025observation}. These developments motivate a unified treatment that connects topological characterization, pump construction, and transport-based measurement.
	
Thouless pumping in lower-dimensional systems provides a practical platform for realizing Chern-insulator models that would otherwise require complex 2D lattice geometries. Replacing one crystal momentum by time, a two-dimensional Bloch Hamiltonian can be converted into a 1D Hamiltonian while preserving the Chern topology on a synthetic torus\cite{celi2014synthetic,ozawa2016synthetic, price2015four}. An explicit mapping would therefore make the Haldane Chern number accessible through quantized pumping in a driven 1D chain. In this work, we develop this program for an extended SSH model with NNN hoppings. Topological pumping cycles are constructed. We visualize the process by tracking the Bloch-vector loop during the modulation and relate the pumped charge to a linking pattern in Bloch space. Reading out the Chern number as quantized 1D transport, we demonstrate that the driven SSH chain inherits the topological features of the Haldane model. 
 
  The remainder of this paper is organized as follows. In Sec.~\ref{sec_model}, we introduce the extended SSH Hamiltonian and its topological characterization. In Sec.~\ref{sec_pump_scheme} we construct a Thouless pumping scheme and present its graphical interpretation in Bloch space. Section \ref{sec_pumping} examines quantized transport in a filled fermionic band and Bloch-oscillation-assisted pumping of a bosonic wave packet. In Sec.~\ref{sec_mapping} we establish the mapping to the Haldane model and show how its Chern number is extracted from one-dimensional transport. Finally, we summarize our findings and discusses possible extensions in Sec.~\ref{sec:summary}.
	
\begin{figure}[htbp]
	\centering
	\includegraphics[width=0.9\columnwidth]{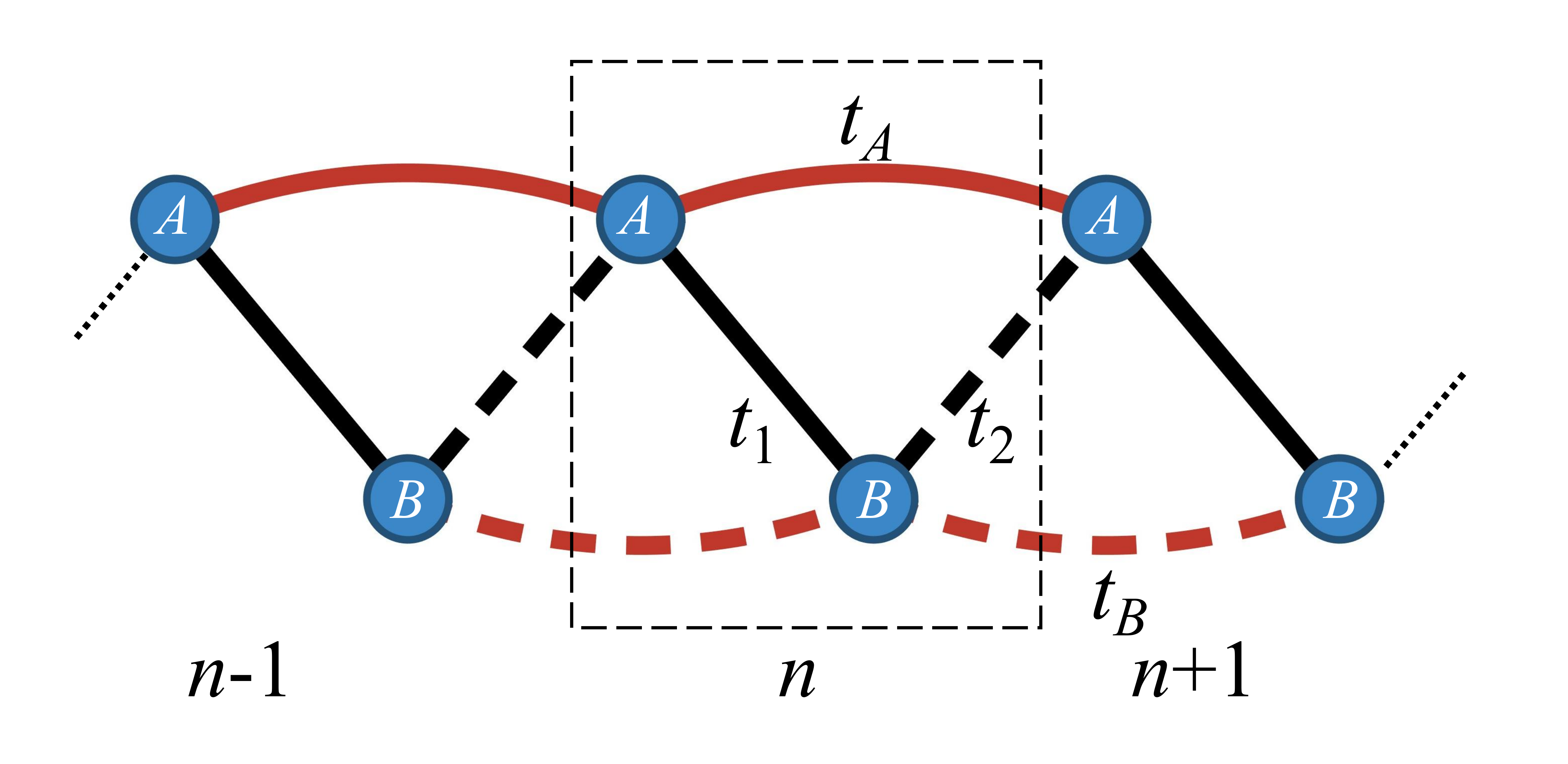}
	\caption{\label{fig:fig1}Schematic diagram of the extended SSH model. Each unit cell (dashed box) contains two sites labeled $A$ and $B$. $t_1$ and $t_2$ represent the intra-cell and inter-cell NN hoppings. NNN hoppings within each sublattice are labeled by $t_A$ and $t_B$. }
\end{figure}

\section{extended SSH model with NNN hopping}\label{sec_model} 
We consider an extended Su-Schrieffer-Heeger (SSH) model \cite{su1979solitons,li2014topological} defined on a one-dimensional lattice with two sublattices ($A$ and $B$ per unit cell). The system is governed by a tight-binding model consisting of staggered nearest-neighbor (NN) hopping and NNN hopping, as illustrated in Fig. 1. The Hamiltonian is given by $H_0 = H_{\mathrm{SSH}} + H_{\mathrm{NNN}}$, where
\begin{eqnarray}
	H_{\mathrm{SSH}} &=& \sum_{n} \left[ t_1 a_{A,n}^{\dagger} a_{B,n} + t_2 a_{A,n+1}^{\dagger} a_{B,n} + \mathrm{H.c.} \right], \label{eq:ssh} \label{eq_SSH}\\
	H_{\mathrm{NNN}} &=& \sum_{n} \left[ t_A a_{A,n}^{\dagger} a_{A,n+1} + t_{B} a_{B,n}^{\dagger} a_{B,n+1} + \mathrm{H.c.} \right]. \label{eq_NNN}
\end{eqnarray}
Here, $a_{\sigma,n}$ ($a_{\sigma,n}^{\dagger}$) denotes the fermionic or bosonic annihilation (creation) operator acting on sublattice $\sigma \in \{A,B\}$ in the $n$-th unit cell. The parameters $t_1$ and $t_2$ represent the intracell and intercell NN hopping amplitudes, respectively. $t_A$ and $t_B$ characterize the NNN hopping strengths. 

Under periodic boundary conditions, a Fourier transformation brings the Hamiltonian to the two-band Bloch form $H = \sum_k \psi_k^{\dagger} H(k) \psi_k$, where $\psi_k^{\dagger} = (a_{A,k}^{\dagger}, a_{B,k}^{\dagger})$. The momentum-space Hamiltonian reads
\begin{equation}\label{eq_Hk}
	H(k) = \begin{pmatrix}
		2t_A \cos k & t_1 + t_2 e^{-ik} \\
		t_1 + t_2 e^{ik} & 2t_B \cos k
	\end{pmatrix}.
\end{equation}
Here $H(k)$ can be decomposed as
\begin{equation} \label{eq:Hk}
	H(k) = \varepsilon_0(k) \mathbf{1} + \bm{h}(k) \cdot \boldsymbol{\sigma},
\end{equation}
with $\mathbf{1}$ the $2\times2$ identity matrix and $\boldsymbol{\sigma}=(\sigma_x,\sigma_y,\sigma_z)$ the Pauli matrices. $\varepsilon_0(k) = (t_A + t_B)\cos k$ represents the diagonal shift and the Bloch vector components are given by
\begin{eqnarray}
	h_x(k) &=& t_1 + t_2\cos k, \nonumber \\
	h_y(k) &=& t_2\sin k,  \\
	h_z(k) &=& (t_A - t_B)\cos k. \nonumber\label{eq:h}
\end{eqnarray}
The two-band dispersion reads
\begin{equation}
	E_{\pm}(k) = \varepsilon_0(k) \pm |\bm{h}| 
\end{equation}
with
\begin{equation}
	|\mathbf{h}| = \sqrt{(t_A - t_B)^2\cos^2 k + t_1^2 + t_2^2 + 2t_1 t_2\cos k}. \label{eq:Delta}
\end{equation}

Equation (\ref{eq:Hk}) describes the action of a magnetic field $\bm{h}(k)$ on a pesudospin in auxiliary space in addition to a scalar energy shift. The topological feature of the system is essentially determined by the chiral property of $\bm{h}(k)$. Note that the standard SSH Bloch Hamiltonian is recovered in the limit $t_A=t_B$. In this case, the bulk Hamiltonian has reflection symmetry and time-reversal symmetry, which ensures $h_z=0$ and the winding of $\bm{h}(k)$ is restricted to a two-dimensional manifold. Therefore, the mapping from the first Brillouin zone (BZ) to the circle traced by $\bm{h}(k)$ ($S^1\to S^1$) is classified by the first homotopy group $\pi_1(S^1)=\mathbb Z$, whose integer index is precisely the winding number ($w=\varphi_\text{Zak}/\pi$ with $\varphi_\text{Zak}$ the Zak phase). Alternatively, the winding number also counts the number of topological defect (the origin) encircled by the endpoints of $\bm{h}$ ($\bm{h}$ circle). A topological phase transition occurs at $t_1=t_2$, where the $\bm{h}$ circle crosses the origin, giving rise to a band crossing at $k=\pm\pi$.

The sublattice-resolved NNN hopping ($t_A\neq t_B$) breaks the reflection symmetry, leading to a nonzero $h_z$-term that changes the Altland-Zirnbauer class from BDI to AI. The topological feature is trivial since $\pi_1(S^2)=0$. However, by introducing a periodic driving in the spirit of Thouless pumping, the 1D BZ and the time dimension form a 2D torus $T^2$. The topology is then characterized by the Chern number $C$, corresponding to the homotopy group $\pi_2(S^2)=\mathbb{Z}$. The Chern number is protected by the band gap instead of chiral symmetry. To construct a pumping scheme, an intuitive thought is to construct a continuous route that connect two topological distinct regimes in the $S^1\to S^1$ mapping without closing the gap. In the following, we construct a pumping scheme, and demonstrate that the Chern number (and hence the quantized charge pump) can be visualized graphically from the Bloch Hamiltonian. 

\begin{figure}[htbp]
	\centering
	\includegraphics[width=0.9\columnwidth]{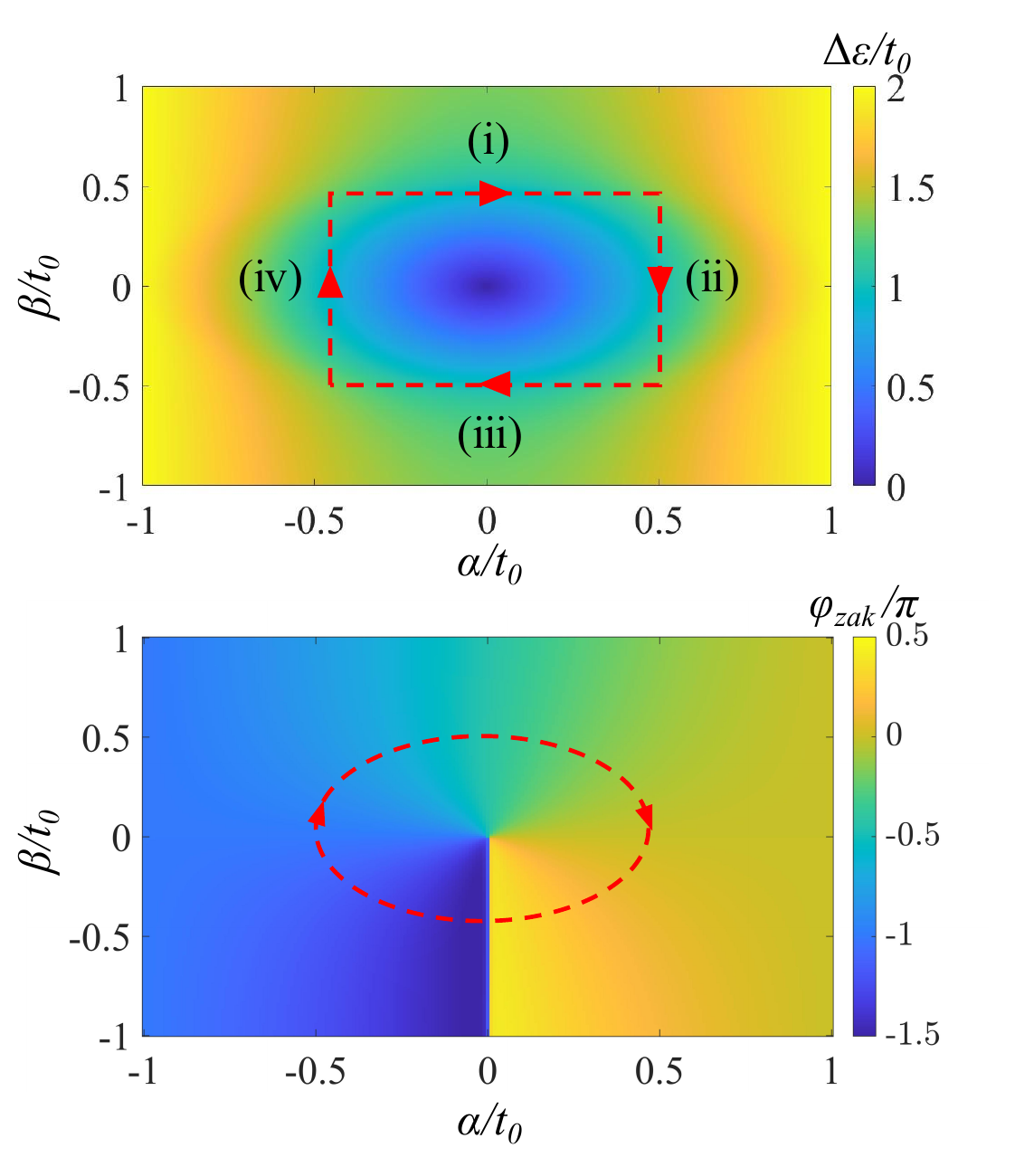}
	\caption{\label{fig:fig2} Parameter space spanned by $\alpha$ and $\beta$. (a) Band gap $\Delta \min(E_+) - \max(E_-)$ as a function of $\alpha$ and $\beta$. The four paths (i)-(iv) constitute a closed loop that encircles the gap closing point. (b) Zak phase $\phi_{\mathrm{Zak}}$ with respect to $\alpha$ and $\beta$. The closed path indicates a typical modulation cycle for Thouless pumping.}
\end{figure}

\section{Thouless pumping scheme}\label{sec_pump_scheme} 
We first set $t_0=t_1+t_2$ as the unit of energy and define $\alpha=t_1-t_2$ as the relative magnitude of staggered hopping rates. The NNN hoppings are reparameterized by $g=t_A+t_B$ and $\beta=t_A-t_B$. We take $\alpha$ and $\beta$ as the driving parameters. Figure 2(a) depicts the band gap $\Delta \varepsilon =\min(E_+)-\max(E_-)$ with respect to $\alpha$ and $\beta$. The Zak phase calculated by 
\begin{equation}
	\varphi_\text{Zak}=i\int_{-\pi}^\pi \langle u_k|\partial_k|u_k\rangle dk.
\end{equation}
is shown in Fig. 2(b). Here $|u_k\rangle$ is the lower band eigenstate of $H(k)$. Along $\beta=0$, one obtains a quantized Zak phase: $\varphi_\text{Zak}=\pi$ for $\alpha<0$ and $\varphi_\text{Zak}=0$ for $\alpha>0$. Nonzero $\beta$ shifts the Bloch vector out of the $h_x$-$h_y$ plane, thus provides a route to connects the two regimes. As $\varphi_\text{Zak}$ varies with parameters in a modulation cycle, a quantized Chern number arises as 
\begin{equation}\label{eq_C}
	C=\frac{1}{2\pi}\int_0^T \partial_t \varphi_\text{Zak}dt.
\end{equation}
To illustrate how the $\bm h$ circle changes in a pumping protocol, we consider a closed path labeled by (i)-(iv) in Fig. 2(a). The trajectory of the $\bm h$ circle for each process is shown in Fig. 3. In process (i), $\alpha/t_0=-0.5\to 0.5$ and $\beta/t_0=0.5$, the $\bm h$ circle shrinks and crosses the $h_z$-axis at $h_z=-\beta$. In process (ii), the $\bm h$ circle rotates (in the $h_x>0$ territory) around the line $(h_x,h_z)=(t_1,0)$. It is also rescaled such that the projection onto the $h_x$-$h_y$ plane remains unchanged. Subsequently, as symmetric to process (i), the $\bm h$ circle grows and crosses the $h_z$-axis at $h_z=\beta$. During (iv), the circle encompasses the origin at the moment with $\beta=0$, and then returns to its initial position by performing a backward rotation in comparison to (ii). The topological essence of the complete pumping process resides in whether the parameter trajectory encircles a bulk gap-closing point in the $(\alpha, \beta)$ plane. Provided that the trajectory winds around this singular point, the Zak phase accumulated over a complete cycle undergoes a $2\pi$ shift, yielding a quantized Chern number $C = 1$ according to Eq. (\ref{eq_C}).

\begin{figure}[tbp]
	\centering
	\includegraphics[width=1\columnwidth]{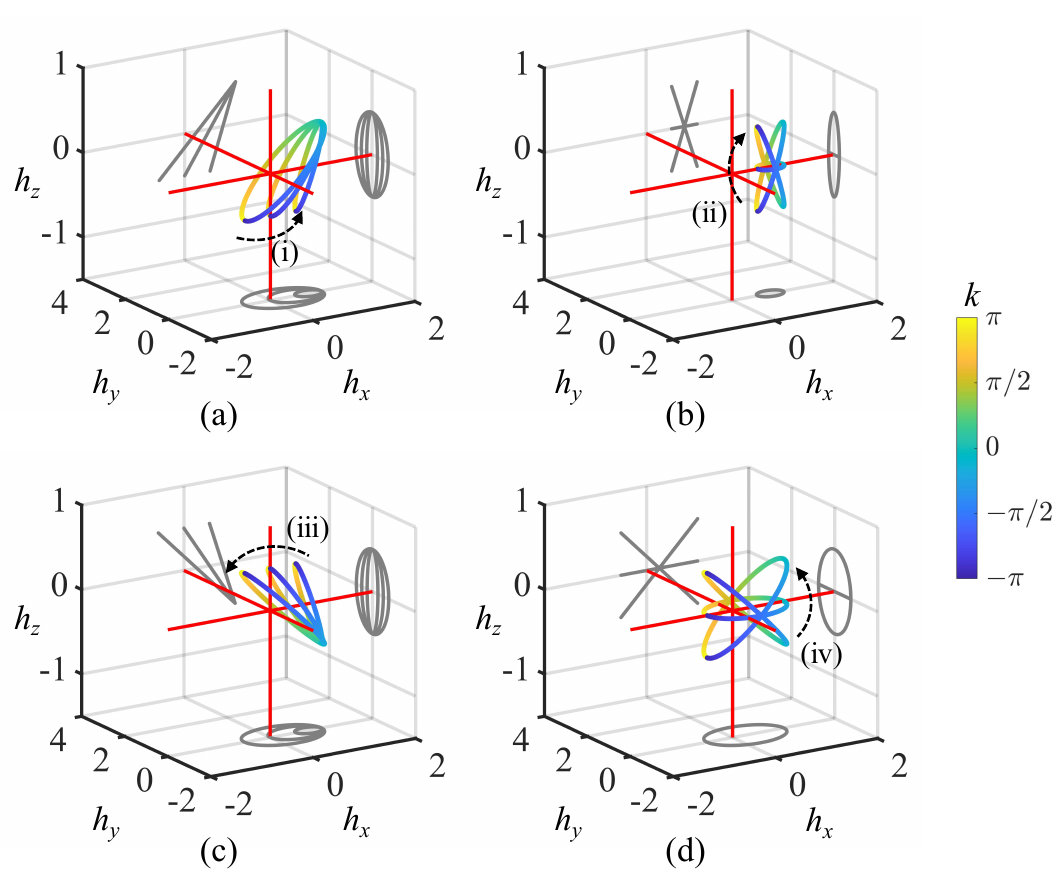}
	\caption{\label{fig:fig3} Variation of the $\bm{h}$ circle for paths (i)-(iv) shown in Fig.~2. The color highlights the value of $k$. Arrows indicate the direction of the trajectory. Projections onto the coordinate planes are shown on the bottom, rear, and left sides.}
\end{figure}

\begin{figure}[tbp]
	\centering
	\includegraphics[width=0.9\columnwidth]{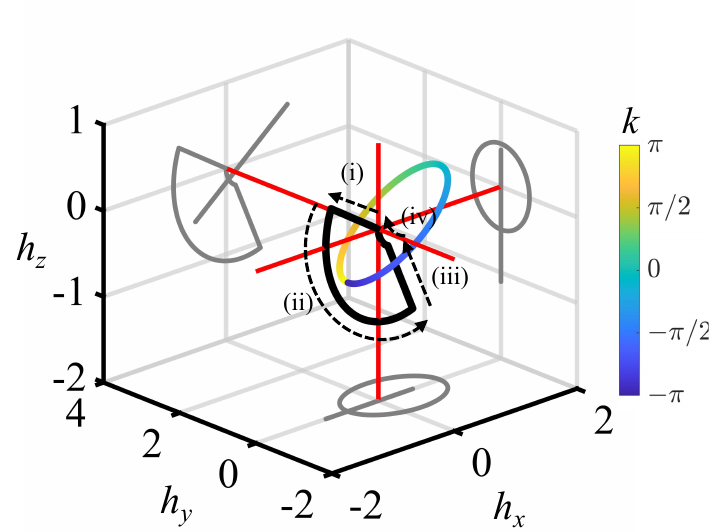}
	\caption{\label{fig:fig4}Linking pattern between the initial $\bm{h}$ circle and the relative trajectory of the origin corresponding to the closed paths (i)-(iv). }
\end{figure}

Instead of calculating the Zak phase and Chern number, figure 3 demonstrate how the $\bm h$ circle wraps the origin once during a period of the modulation. One can also interpret this behavior as the origin passes through the $\bm h$ circle once. Such a graphical representation provides a practical way to construct Thouless pumping schemes\cite{kraus2012topological,ke2016topological, grinberg2020robust,xia2021experimental, tangpanitanon2016topological,jurgensen2021quantized,mostaan2022quantized}. To illustrate the topological feature, we hold the initial circle fixed in Bloch space, and regard its motion as a relative movement of the origin. The result is shown in Fig. 4, exhibiting a typical linking pattern (with linking number 1) that is in accordance with $C=1$. 

\section{Observation of quantized pumping} \label{sec_pumping}
The quantized Chern number underpins the robust quantized transport in the adiabatic regime. In real-time evolution, the pumped charge can be obtained by integrating the instantaneous current density, \textit{i.e.}, $\Delta n(t) = \int_{0}^{t} j(t')\, dt'$. For a translationally invariant system with the lower band fully occupied by spinless fermions, the current density is obtained by the averaging the expectation value of velocity operator over Brillouin zone:
\begin{equation}
	j(t) = \frac{1}{2\pi} \int_{\mathrm{BZ}} \langle \psi_k(t) | \partial H_k(t)/\partial (\hbar k) | \psi_k(t) \rangle \, dk. \label{eq:current}
\end{equation}
Here $|\psi_k(t)\rangle$ represents the time-evolved state starting from the initial Bloch state $|u_k\rangle$, which can be numerically obtained by solving the time-dependent Schr\"odinger equation under a specific pumping protocol. 

\begin{figure}[tbp]
	\centering
	\includegraphics[width=0.9\columnwidth]{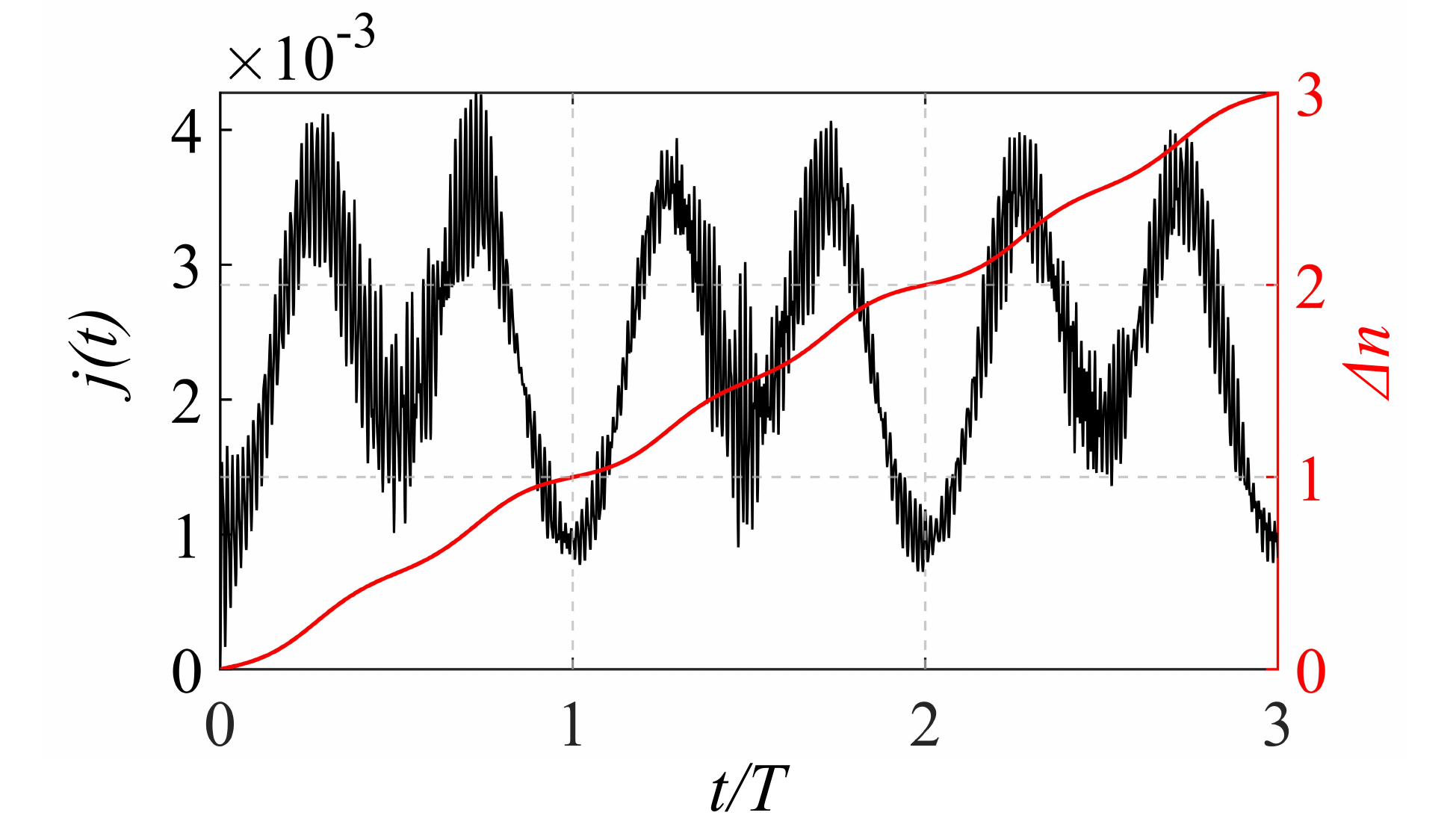}
	\caption{\label{fig:fig5}Evolution of the charge current $j(t)$ and the pumped charge $\Delta n$ for the modulation scheme given in Fig.~2(b).}
\end{figure}
	
We take a modulation scheme with parameters following the route shown in Fig. 2(b), specifically, $\alpha=A\cos(\omega t)$ and $\beta=A\sin(\omega t)$ with $T=2\pi/\omega$ the period. We have set $A/t_0=0.5$ and $\hbar\omega/t_0=0.05$. The evolution of the instantaneous current density $j$ and the pumped charge $\Delta n$ as a function of time are plotted in Fig. 5. As the system adjusts to the varying parameters, its response to the initial conditions may give rise to high-frequency oscillations of $j(t)$, which can be different in each modulation cycle. However, the pumped charge within each complete cycle is quantized to Chern number ($C=1$) of the occupied band.

Thouless pumping for fermions relies on an initially occupied band. For bosonic wave packets, however, such a uniform band occupation is not naturally available. We therefore turn to a Bloch-oscillation-assisted protocol, in which a weak linear gradient is added to the driven lattice. The Hamiltonian is given by
\begin{equation}
	H(t)=H_0(t)+F\sum_j jn_j,
\end{equation}
where $F=\hbar\omega_F$ represents the detuning between two neighboring sites. $n_j=a_j^\dag a_j$ with $j$ the site index. Here $a_j$ is connected to the annihilation operators in Eqs. (\ref{eq_SSH}) and (\ref{eq_NNN}) through $a_{A,n}=a_{2n-1}$ and $a_{B,n}=a_{2n}$. Upon a gauge transformation $a_j^\dag=e^{-i\omega_Fjt}b_j^\dag$, the tilt appears as a time-dependent Peierls phase in the hopping terms. Equivalently, the quasimomentum of a wave packet is swept through the Brillouin zone as $k(t)=k_0-\omega_F t$. Thus, the same time variable that drives the pumping cycle also enforces a dynamical sampling of momentum space through Bloch oscillations. This mechanism is particularly useful for bosonic implementations because it replaces the stringent requirement of preparing a uniformly filled band by a controlled real-time scan of the band geometry\cite{ke2017multiparticle,haug2020topological}. 

For an initial Gaussian wave packet prepared at $k=k_0$, the transport is characterized by the center-of-mass displacement $\Delta X(t)=X(t)-X(0)$ (with $X(t)=\sum_jj|\psi_j(t)|^2$), which evolves as 
\begin{equation}
	\Delta X(t)=\int_0^t v_g(k_0,t')dt'.
\end{equation} 
In the adiabatic regime, the semiclassical velocity is given by
\begin{equation}
	v_g(k_0,t)=\frac{1}{\hbar}\frac{\partial\varepsilon_n(k_0-\omega_F t,t)}{\partial k} +\mathcal F_n(k_0-\omega_F t,t),
\end{equation}
where $\varepsilon_n(k,t)$ is the instantaneous energy of the $n$-th band and $\mathcal{F}_n(k,t)=i(\langle\partial_t u_k|\partial_k u_k\rangle-\langle\partial_k u_k|\partial_t u_k\rangle)$ is the Berry curvature defined in the case of $F=0$. As proposed in refs. [ke2020topological]\cite{ke2020topological}, when $\omega_F/\omega=p/q$, with $p$ and $q$ coprime and $T=2\pi/\omega$, the net drift over the common period $qT$ is given by
\begin{eqnarray}
	 \Delta X(qT)/d &=& \frac{1}{d}\int_0^{qT}\mathcal F_n(k_0-\omega_F t,t)dt\nonumber\\
	 &\simeq& \frac{1}{2\pi}\int_0^{2\pi/d}\int_0^{qT} \mathcal F_n(k,t)dt\nonumber\\
	 &=& qC_n. 
\end{eqnarray}
Here $d=2$ counts the size of each unit cell and $C_n$ is the Chern number for the corresponding band. The dispersion velocity averages out over the closed Bloch-pumping trajectory. This result relies on the approximation that the first integral can be independent of the value of $k_0$ as the BZ is uniformly sampled by Bloch oscillation. Consequently, the measured displacement per common period $qT$ provides a one-dimensional, time-domain readout of the band Chern number.

\begin{figure}[htbp]
	\centering
	\includegraphics[width=0.9\columnwidth]{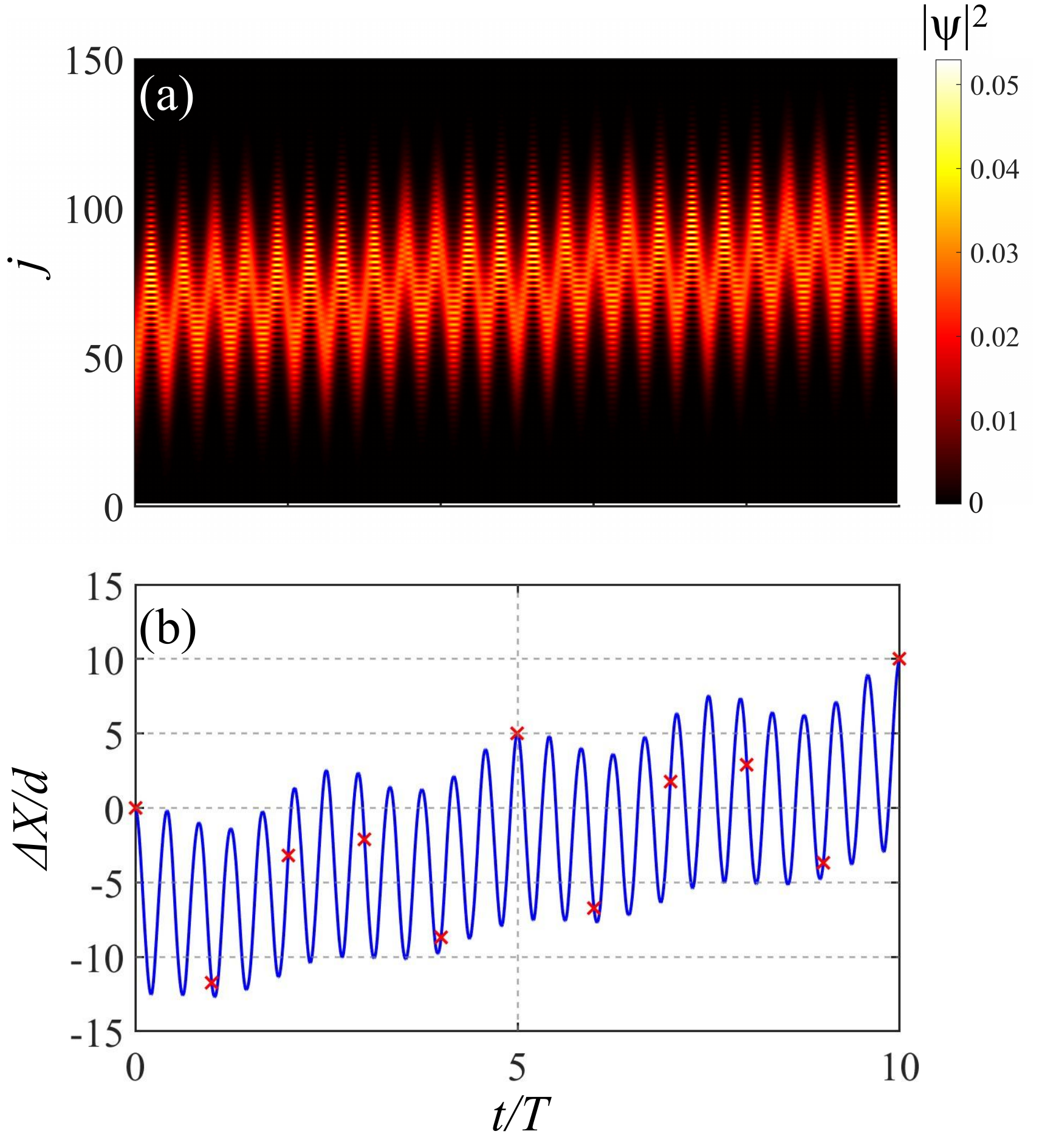}
	\caption{\label{fig:fig6} Oscillation for an initial Gaussian wave packet under the combined effect of a gradient potential and pumping modulation. (a) Density evolution over two modulation periods. (b) Corresponding center-of-mass displacement as a function of time. The red crosses denote the displacement at multiple complete driving cycles. The parameters are set to $\omega_F/\omega = 6/5$, $d = 2$, $\sigma = 25$, $j_0 = 50$, and $k_0 = 0$.}
\end{figure}
	
We simulate this protocol by employing a Gaussion wave packet in the lower band\cite{ke2020topological},
\begin{equation}
	\psi_j(0)=\mathcal N\exp\!\left[-\frac{(j-j_0)^2}{4\lambda^2}\right]u_{-,j}(k_0,0)
\end{equation}
where $\mathcal N$ is the normalizing factor, $u_{-,j}(k_0)$ is the instantaneous lower-band spinor at $t=0$. Here, $u_{-,j}(k_0,0)=1/\sqrt{2}$ for odd $j$ and $u_{-,j}(k_0,0)=-1/\sqrt{2}$ for even $j$. We have set $k_0=0$, $\lambda=25$, $j_0=50$ and $F=0.06$. We adopt the same modulation scheme as in the previous case [see Fig. 2(b)] and set $\omega_F/\omega=6/5$. As shown in Fig. 6 (a), the bosonic wave packet undergoes Bloch-like oscillations during the pump. The center-of-mass motion of the density profile is depicted in Fig. 6(b). The red crosses mark the selected positions used to extract the net drift. Despite the large intra-cycle motion, we observe $\Delta X/d=5$ and $10$ respectively for $t=5T_m$ and $10T_m$ as expected, revealing the accumulated topological displacement. 

\section{Mapping to the Haldane model}\label{sec_mapping}
The periodic driving discussed above shows that the extended SSH chain can carry a Chern number on the synthetic torus formed by momentum and time\cite{mancini2015observation,livi2016synthetic,an2017direct,fabre2022laughlin}. We now take this strategy to connect the driven 1D system to the Haldane model. The mapping provides both an intuitive definition and a practical measurement strategy for the topological invariant. The Chern number of a two-dimensional model can be equivalently extracted from the quantized charge transport in one-dimensional adiabatic cycles. 

We start by briefly reviewing the Haldane model on the honeycomb lattice \cite{haldane1988model,jotzu2014experimental},
\begin{eqnarray}
	H &=& \sum_i M_i c_i^\dagger c_i
	+ \tau_1 \sum_{\langle i,j\rangle} c_i^\dagger c_j
	+ \tau_2 \sum_{\langle\langle i,j\rangle\rangle}
	e^{i\phi_{ij}} c_i^\dagger c_j .
\end{eqnarray}
Here $M_i=\pm M$ is the staggered sublattice potential, $\tau_1$ and $\tau_2 e^{i\phi_{ij}}$ are the nearest-neighbor and NNN hopping rates, respectively. The phases $\phi_{ij}=\pm\phi$ break time-reversal symmetry while keeping zero net magnetic flux through the lattice. In a two-sublattice Bloch basis, and after choosing the momentum coordinates shown in Fig. 7, the momentum-space Hamiltonian can be written as
\begin{eqnarray}\label{eq_Haldane_Hk}
	H_k^{\rm Hal} &=&
	\begin{pmatrix}
		2S_+ \cos \frac{3k_x}{2} + U_+ & \tau_1 + Re^{-i\frac{3k_x}{2}} \\
		\tau_1 +R e^{i\frac{3k_x}{2}} & 2S_- \cos \frac{3k_x}{2} + U_-
	\end{pmatrix},
\end{eqnarray}
with
\begin{eqnarray}
	S_\pm &=& 2\tau_2\cos\left(\frac{\sqrt{3}k_y}{2}\pm\phi\right), \nonumber \\
	R &=& 2\tau_1\cos\left(\frac{\sqrt{3}k_y}{2}\right), \\
	U_\pm &=& \pm M + 2\tau_2\cos\!\left(\sqrt{3}k_y\mp\phi\right).\nonumber
\end{eqnarray}
This form separates the two roles played by the two crystal momenta $k_x$ and $k_y$. This observation leads directly to a driven extended SSH realization. We identify the momentum components of the Haldane model with the physical momentum and the pumping time coordinate of the one-dimensional chain as
\begin{equation}
	\frac{3}{2}k_x=k,\qquad
	\frac{\sqrt{3}}{2}k_y =\omega t,
\end{equation}
where $k$ is the 1D momentum and $\omega$ is the driving frequency. The corresponding one-dimensional Hamiltonian is $H_{1D}(t)=H_0(t)+H_\text{on-site}(t)$ with additional on-site potentials given by
\begin{equation}
	H_\text{on-site} = \sum_n \left[\mu_A(t)a_{A,n}^\dagger a_{A,n}
	+\mu_B(t)a_{B,n}^\dagger a_{B,n}\right].
\end{equation}
The time-dependent Bloch Hamiltonian yields 
\begin{eqnarray}\label{eq_ext_ssh}
	H_{\rm ext}(k,t) &=&
	\begin{pmatrix}
		2t_A\cos k+\mu_A &
		t_1+t_2 e^{-ik} \\
		t_1+t_2 e^{ik} &
		2t_B\cos k+\mu_B
	\end{pmatrix},
\end{eqnarray}
which has the same form as Hamiltonian (\ref{eq_Haldane_Hk}). The equality between the two models is obtained by choosing the parameters listed in Table \ref{tab_mapping}.
\begin{table}[htbp]
	\centering
	\caption{\label{tab_mapping} Mapping between the Haldane model and the driven extended SSH model.}
	\begin{ruledtabular}
		\begin{tabular}{c|c|c}
			\textbf{Haldane} & \textbf{Extended SSH} & \textbf{Expression} \\
			\hline
			$\dfrac{3}{2}k_x$ & $k$ & physical momentum \\
			\hline
			$\dfrac{\sqrt{3}}{2}k_y$ & $\omega t$ & time coordinate\\
			\hline
			$\tau_1$ & $t_1$ & $\tau_1$ \\
			\hline
			$R$ & $t_2$ &$2\tau_1\cos (\omega t)$ \\
			\hline
			$S_+$ &$t_A$ & $2\tau_2\cos(\omega t + \phi)$\\
			\hline
			$S_-$ & $t_B$ & $2\tau_2\cos(\omega t - \phi)$\\
			\hline
			$U_+$ & $\mu_A$ & $M + 2\tau_2\cos(2\omega t - \phi)$ \\
			\hline
			$U_-$ & $\mu_B$ & $-M + 2\tau_2\cos(2\omega t + \phi)$ \\
		\end{tabular}
	\end{ruledtabular}
\end{table}

\begin{figure}[htbp]
	\centering
	\includegraphics[width=0.9\columnwidth]{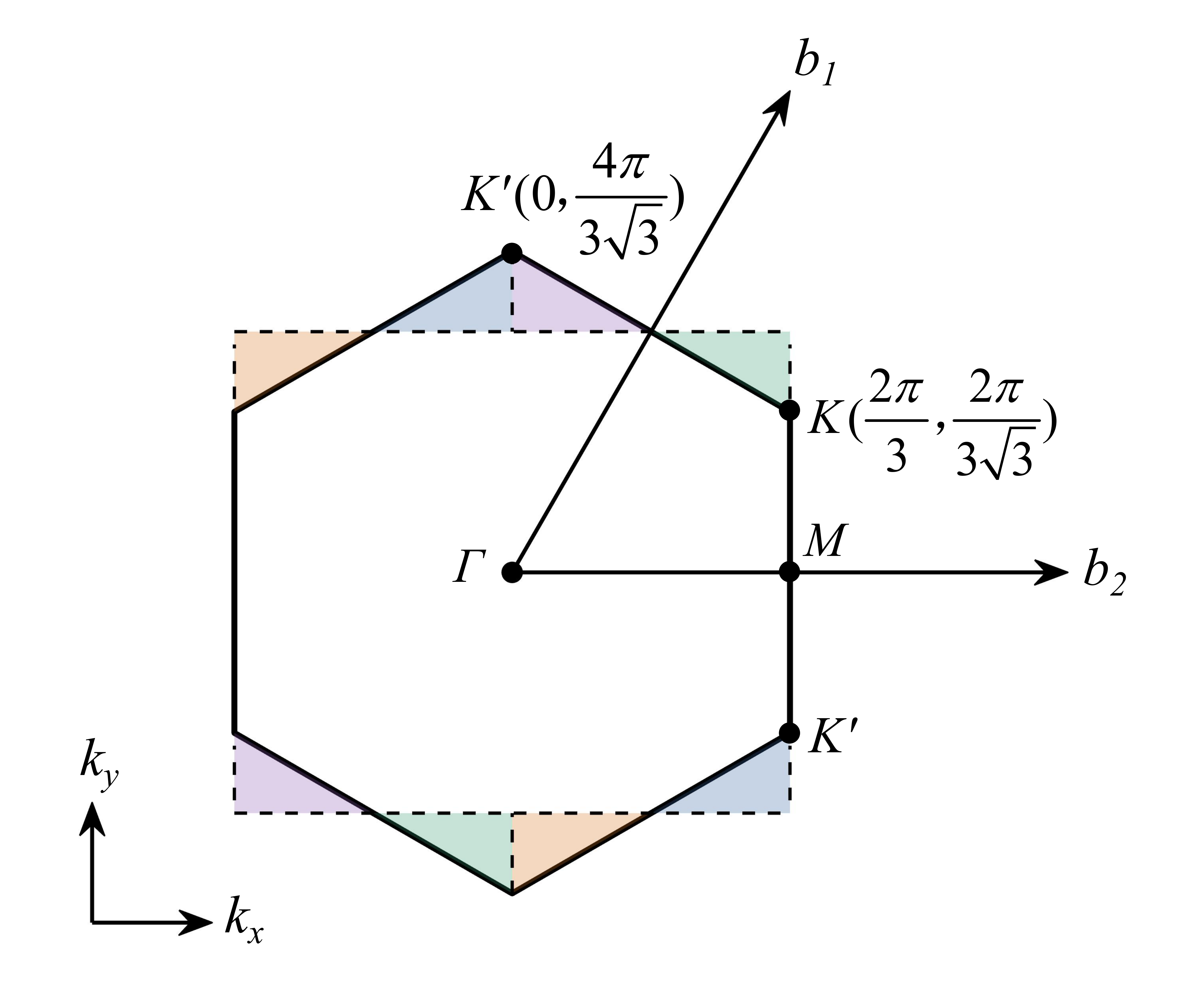}
	\caption{\label{fig:fig7}First BZ of the Haldane honeycomb lattice (area denoted by solid lines). Reciprocal lattice vectors are $\bm{b}_1=2\pi(1/3,1/\sqrt 3)$ and  $\bm{b}_2=2\pi(2/3,0)$. We have set the lattice constant $a=1$. High-symmetry points $\Gamma$, $M$, $K$, and $K^{\prime}$ are labeled in the plot. The BZ can be rearranged into a rectangle indicated by the dashed lines. }
\end{figure}
	
The Chern number of the Haldane model can be obtained by integrating the Berry curvature $\mathcal{F}_n(k_x,k_y)$ over the hexagonal BZ. To connect the Chern number to the pumped charge in the 1D driving scheme, we perform a shear transformation for the BZ to obtain a rectangle of equal area, as shown in Fig. 7. The ranges of $k_x$ and $k_y$ are then given by $k_x\in[-2\pi/3,2\pi/3)$, $k_y\in[-\pi/\sqrt3,\pi/\sqrt3)$, giving rise to $k\in[-\pi,\pi)$ and $\omega t\in[-\pi/2,\pi/2)$. However, the driven 1D model (\ref{eq_ext_ssh}) has a period of $2\pi$. For a complete modulation cycle, the pumped charge is two times the Chern number since the BZ has been covered twice. This stems from the peculiar boundary properties of the hexagonal BZ, \textit{i.e.}, $k$ is shifted by $\pi$ ($k\mapsto k+\pi$) for $\omega t=-\pi/2\to\pi/2$. 

Again, we decompose the Bloch Hamiltonian as 
\begin{equation}
	H_{\rm ext}(k,t)=\varepsilon_0(k,t)\mathbf{1}
	+\bm{h}(k,t)\cdot\boldsymbol{\sigma},
\end{equation}
where the scalar part $\varepsilon_0=(t_A+t_B)\cos k+(\mu_A+\mu_B)/2$ only shifts the band energies. The $\bm h$ circle governed by
\begin{eqnarray}
	h_x(k,t) &=& t_1+t_2(t)\cos k,\nonumber\\
	h_y(k,t) &=& t_2(t)\sin k,\\
	h_z(k,t) &=& [t_A(t)-t_B(t)]\cos k
	+\frac{\mu_A(t)-\mu_B(t)}{2}.\nonumber
\end{eqnarray}
varies in 3D Bloch space, which determines the topological feature. Compared with the pump constructed in Sec.~\ref{sec_pump_scheme}, the present modulation is fixed by the mapping instead of choosing an arbitrary loop in parameter space. Evolving the system in time $t$ is equivalent to scanning the Haldane momentum $k_y$. The graphical representation of a nontrivial pump with $C=1$ is illustrated in Fig.~8. For each fixed $t$, the quasi-momentum $k$ traces a closed $\bm{h}$ loop. During a half period, the $\bm{h}$ circle wraps around the origin once and returns to its initial position, up to a $\pi$ shift in $k$.

\begin{figure}[htbp]
	\centering
	\includegraphics[width=0.9\columnwidth]{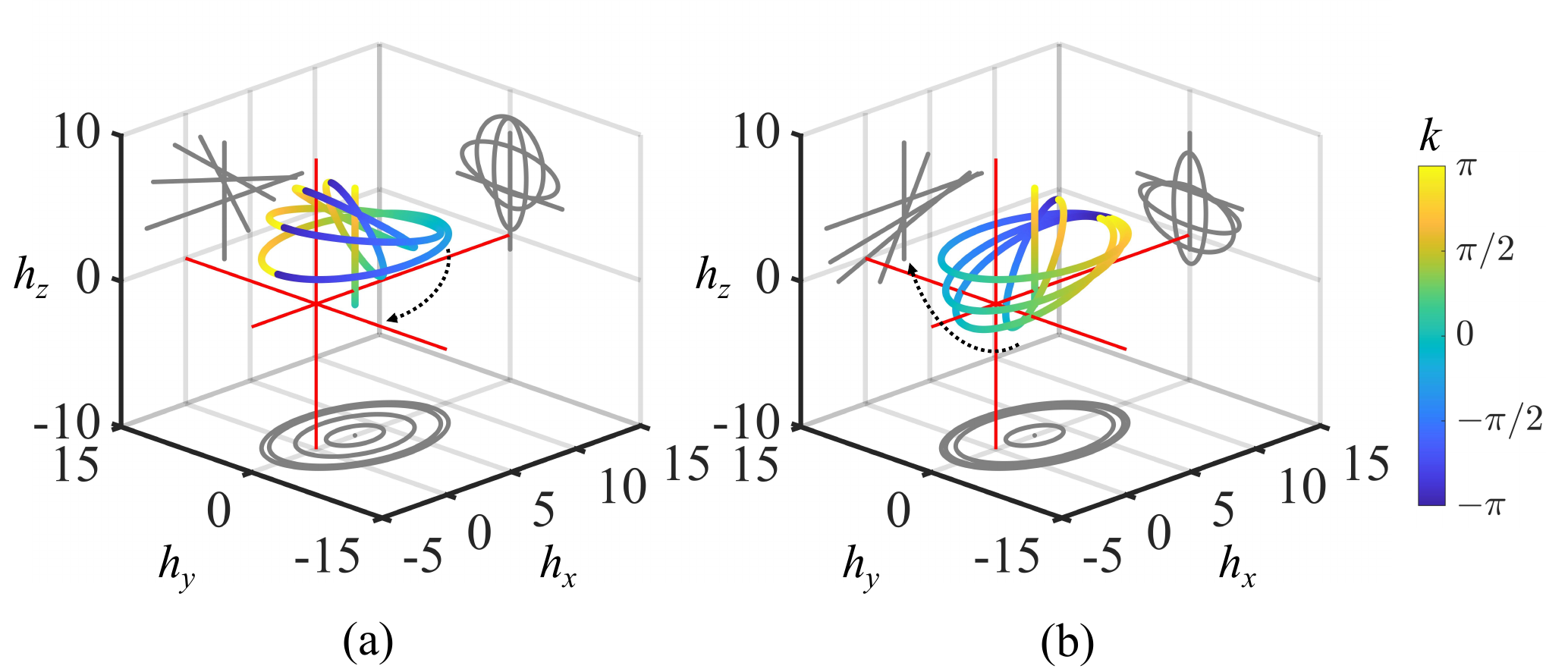}
	\caption{Graphical representation of the pumping scheme corresponding to the Haldane model. The $\bm h$-circle is varied with respect to $t$, with the color highlighting the value of $k$. The arrow shows the direction in which the circle moves. Projections onto a plane are shown on the bottom, rear, and right sides. We have set $\tau_1/\tau_2=3$, $M=3$ and $\phi=0.5\pi$.}
\end{figure}

\begin{figure}[tbp]
	\centering
	\includegraphics[width=0.9\columnwidth]{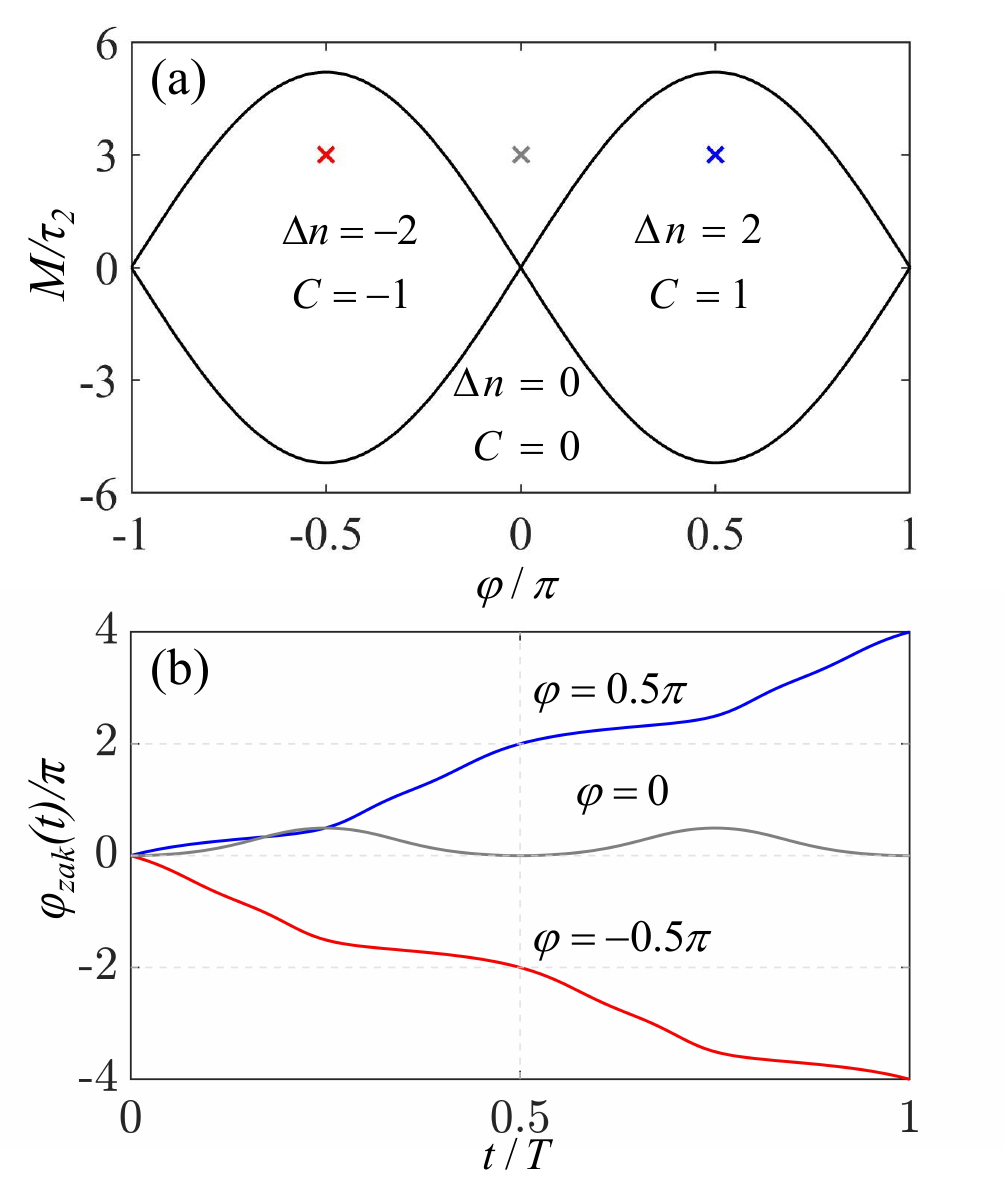}
	\caption{\label{fig9}(a) Topological phase diagram for the driven extended SSH model, with respect to the Peierls phase $\phi$ and the detuning $M$. The three phases indicate the parameter regions that yield $\Delta n=2$, $-2$ and $0$ for a complete modulation cycle. They correspond to the $C=1$, $-1$ and $0$ regimes in the Haldane model. (b) Zak phases versus modulation time $t$ with parameters marked in (a). We have set $\tau_1/\tau_2= 3$}
\end{figure}

The phase diagram obtained from the driven SSH Hamiltonian reproduces the Haldane model. Figure 9(a) shows the resulting domains with $C=1$, $C=0$, and $C=-1$. The sign of the Chern number is determined by the pumping direction. Gap closings occur when $\mathbf{h}(k,t)=0$, which gives the familiar critical lines $M=\pm 3\sqrt{3}\tau_2\sin\phi$. The Zak-phase evolution in Fig.~9(b) gives the same information in 1D language. As the system is varied adiabatically, the Zak phase $\varphi_\text{Zak}$ accumulates $4\pi C$ during a full period. Turning to the transport properties, for a uniformly occupied lower band, the pumped charge is 
\begin{equation}
	\Delta n(T)=\int_0^T j(t)\,dt=2C
\end{equation}
with $T=2\pi/\omega$. This is verified in Fig.~10, where we have set $M=3$ and $\phi=0.5\pi $ corresponding to $C=1$. We observe quantized charge pumping with $\Delta n=2$ during each complete modulation cycle. 

\begin{figure}[tbp]
	\centering
	\includegraphics[width=0.9\columnwidth]{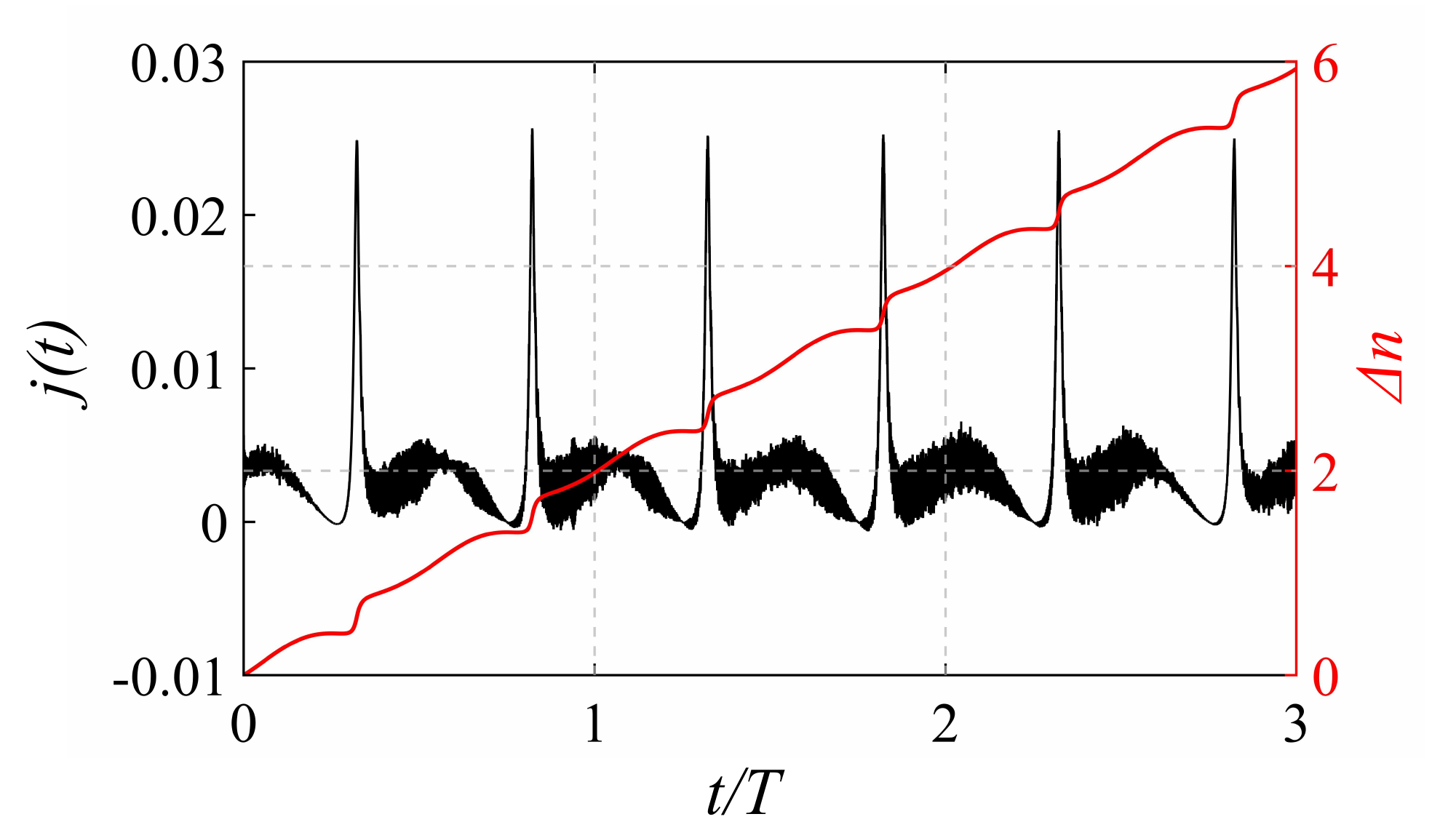}
	\caption{\label{fig:fig10}Evolution of the charge current $j(t)$ and the pumped charge $\Delta n(t)$ for the pump scheme that captures the Haldane-model topology. }
\end{figure}

Finally, the same Chern number can be read out without preparing a filled band by using the Bloch-oscillation-assisted protocol. The weak tilt sweeps the wave packet through $k$ while the drive scans the time coordinate. The evolution samples the same $(k,t)$ parameter space. Starting from a Gaussian wave packet, the stroboscopic displacement therefore satisfies
\begin{equation}
	\frac{\Delta X(qT)}{d}\simeq 2qC, 
\end{equation}
for $\omega_F/\omega=p/q$ in the adiabatic regime. As demonstrated in Fig. 11 with $\omega_F/\omega=4/3$, a drift is superimposed on the Bloch oscillation. The center-of-mass displacement reaches $\Delta X/d=6$ and $12$ at $t=3T$ and $6T$, respectively. The generalized SSH chain therefore provides a 1D dynamical realization of the Haldane Chern band, with its topological invariant accessible either from a filled-band Thouless pump or from the center-of-mass motion of a single bosonic wave packet. 

\begin{figure}[tbp]
	\centering
	\includegraphics[width=0.9\columnwidth]{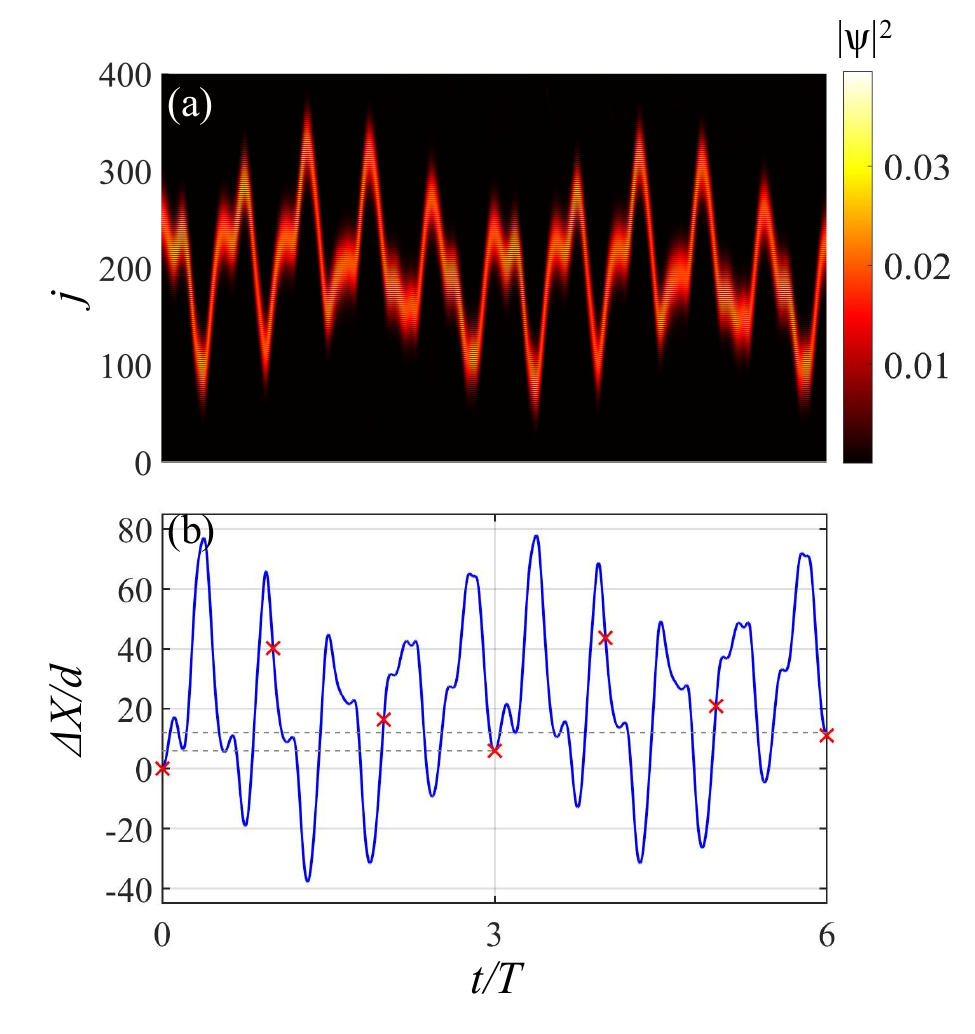}
	\caption{\label{fig:fig11} Quantized drifting oscillation for an initial Gaussian wave packet under the same pumping modulation as in Fig.~10. (a) Real-space density evolution. (b) Center-of-mass displacement with respect to time. The crosses mark the data for multiple complete cycles. The two horizontal dashed lines represent $\Delta X/d=6$ and $12$. Parameters are set to $\sigma = 20$, $k_0 = 0$, $j_0 = 250$, and $\omega_F/\omega = 4/3$.}
\end{figure}

\section{\label{sec:summary} SUMMARY}
In summary, we have studied Thouless pumping in an extended SSH model with NN and NNN hopping modulations. The work is motivated by the simple observation that a one-dimensional adiabatic cycle can serve as a measurable representation of a two-dimensional Chern band. We presented a graphical description in Bloch space, where the pumping process is related to the linking pattern between the Bloch-vector loop and the relative trajectory of the gap-closing point. This representation gives an intuitive way to design pumping paths and to identify the associated Chern number without relying only on the static band topology.

We confirmed the pumping response from two complementary dynamical probes. For a filled band, the integrated current gives a quantized pumped charge. For a bosonic wave packet, Bloch-oscillation-assisted pumping converts the same invariant into a stroboscopic center-of-mass displacement. The latter protocol is particularly useful for cold-atom experiments, where preparing a uniformly occupied band is not always convenient, while applying a weak force and tracking the density profile are standard operations \cite{gross2017quantum,ke2020topological}. The required hopping modulation and synthetic gauge structure may also be implemented in optical superlattices, Raman-assisted tunneling schemes, or photonic lattice platforms \cite{lohse2016thouless,miyake2013realizing,jotzu2014experimental,rechtsman2013photonic}.

We further demonstrated the mapping between the Haldane model and a driven extended SSH model. The first BZ of a 2D Chern-insulator Hamiltonian is replaced by the space spanned by quasimomentum and time of the driven chain. Therefore, the Chern number can be measured through one-dimensional quantized transport rather than through a direct reconstruction of Berry curvature in a 2D BZ. This framework can be extended to other two-band Chern models, such as the Qi-Wu-Zhang model \cite{qi2006topological}, and may provide a useful route for studying interacting, disordered, or higher-order pumping problems in experimentally flexible one-dimensional systems\cite{wauters2019localization,hayward2021effect,marra2020topologically,nakajima2021competition, nakagawa2018breakdown,bertok2022splitting,zeng2016fractional,greschner2020topological,berg2011quantized,kuno2020interaction}.
	
\bibliography{TP.bib}
\end{document}